\documentclass[aps,twocolumn,superscriptaddress]{revtex4}
\usepackage{graphicx}

\begin{document}

\title{Critical scaling of thermal transport in model A dynamics for a
superconductor}

\author{Iddo Ussishkin}
\affiliation{William I. Fine Theoretical Physics Institute,
University of Minnesota, Minneapolis, MN 55455}
\author{David A. Huse}
\author{S. L. Sondhi}
\affiliation{Department of physics, Princeton University,
Princeton, NJ 08544}

\begin{abstract}
We consider the scaling of thermal transport in the presence of
electric and magnetic fields near the finite temperature transition
from the metallic normal state to the superconductor. We do so with
fully relaxational, model A, dynamics for the order parameter and
particle-hole symmetry. This enables us to determine the exact scaling
dimension of the heat current operator and hence critical exponents for
all transport coefficients in terms of the (approximately) known values
of the correlation length exponent and the dynamic scaling exponent. In
particular, we determine the critical behavior for the Nernst
coefficient.
\end{abstract}

\maketitle

\section{Introduction}
\label{introduction}

The discovery of high-temperature superconductors has led to a
renewed interest in superconducting fluctuations. In these
materials, the effect of fluctuations may be strong, with a
relatively large critical regime, due to their short coherence
lengths, quasi-two-dimensionality and high transition
temperatures~\cite{Fisher-Fisher-Huse}. Neglecting fluctuations of
the gauge potential, the transition to the superconducting state
is described by a complex order parameter and is of the XY
universality class (in three dimensions, due to interlayer coupling).
This description holds well in strongly
type-II superconductors except extremely close to the critical
temperature $T_c$ where one crosses over to the critical behavior
of a charged superconductor~\cite{invertedXY}. The description of
dynamic properties, such as transport, requires coupling this
thermodynamic description to a dynamic equation, which for
superconductors is usually assumed to be model A in the
classification of Hohenberg and
Halperin~\cite{Hohenberg-Halperin}.

Of particular interest in this context are the recent measurements
of the Nernst coefficient in the cuprates~\cite{Ong}. The large
Nernst signal observed in these experiments above $T_c$ is likely
the contribution of superconducting fluctuations. This
interpretation provides a quantitative agreement with experiment
in overdoped samples in the regime where a Gaussian approximation
is applicable~\cite{Ussishkin-Sondhi-Huse}, as well as in the
vortex liquid regime~\cite{Mukerjee}. In underdoped samples it
requires a stronger effect of fluctuations, which would have
important implications for the physics of the pseudogap
regime~\cite{Ussishkin-Sondhi}.

In this paper, we focus our attention on the scaling behavior of
thermal transport in the XY critical regime. The transport
coefficient of interest for the Nernst effect is the transverse
thermoelectric response $\alpha_{xy}$~\cite{alpha-xy}. For this,
we consider the scaling of the heat current operator, which also
allows us to treat the thermal conductivity on the same footing.
As with the conductivity~\cite{Fisher-Fisher-Huse,Wickham-Dorsey},
it is necessary to specify the dynamics associated with the order
parameter for analyzing thermal transport. We then find the
scaling dimension of the heat current operator, and obtain general
scaling forms for thermal transport. The critical behavior of both
the thermal and thermoelectric transport coefficients is then
deduced.

We follow the usual assumption for a superconductor, and consider
relaxational dynamics for a non-conserved order parameter, or model A
dynamics~\cite{Hohenberg-Halperin,fn-model}.
We further assume that the model is
particle-hole symmetric which allows much progress to be made.
When needed, the order parameter may be coupled to an
electromagnetic field, whose dynamics we ignore, keeping our
calculation in the XY regime~\cite{Lennart}.

The main result of this paper is simply stated: the heat current
operator in model A,
\begin{equation}\label{heat-current}
\mathbf{j}^Q = - \frac{\partial \psi^*}{\partial t} \nabla \psi +
\text{c.c.} ,
\end{equation}
does not acquire an anomalous dimension. This result is surprising on
two counts: First, the heat current is not a conserved quantity in
model A and second, the scaling dimension of the heat current does not
involve the dynamic critical exponent $z$, despite its dynamic nature.
Technically we will find that there is a ``hidden'' conservation law at
play in the model. We note that a similar absence of an anomalous
dimension is well known for the electric current \cite{Fisher-Fisher-Huse},
\begin{equation}
\mathbf{j} \propto - i \psi^* \nabla \psi + \text{c.c.} .
\end{equation}
In this case, however, there are conserved (solenoidal) supercurrents in
the purely {\it static} theory and thus the result is less mysterious.

With this result, critical exponents of various thermal transport
coefficients may now be deduced. In particular, we find that to linear
order in magnetic field, the transverse thermoelectric response
$\alpha_{xy}$ scales in three dimensions as $\alpha_{xy} / B \propto
\xi^{z - 1}$, where $\xi$ is the (temperature-dependent) correlation
length of the superconductor. Whether this prediction may be observable
in the critical 3D-XY regime of cuprates remains to be seen.

Below we present our calculations in detail:  After defining the
model in Sec.~\ref{modelA}, we obtain the anomalous dimension of
the heat current operator (or rather, its absence thereof) in
Sec.~\ref{dimension}. We briefly address the problem of
introducing a temperature gradient in Sec.~\ref{Tgradient}. The
scaling relations for thermal transport and their consequences are
discussed in Sec.~\ref{transport}, which is followed by a brief
summary.

\section{Model A}
\label{modelA}

In this paper, we consider model A dynamics for a complex
superconducting order parameter. The order parameter dynamics in this
model is given by the stochastic equation
\begin{equation}\label{model-A}
\frac{\partial \psi}{\partial t} = - \Gamma_0 \frac{\delta F}{\delta
\psi^*} + \zeta ,
\end{equation}
with the Landau-Ginzburg-Wilson free energy
\begin{equation}\label{free-energy}
F = \int d \mathbf{x} \left( r_0 |\psi|^2 + u_0 |\psi|^4 + |\nabla
\psi|^2 \right) .
\end{equation}
In Eq.~(\ref{model-A}), $\Gamma_0$ is the relaxation rate for the order
parameter, which we assume in this paper to be real (making the model
particle-hole symmetric). Thermal fluctuations are introduced with
$\zeta$, which is a Gaussian white noise with correlator
\begin{equation}\label{noise}
\langle \zeta^* (\mathbf{x}, t) \zeta (\mathbf{x'}, t') \rangle = 2
\Gamma_0 T \delta (\mathbf{x} - \mathbf{x'}) \delta (t - t') .
\end{equation}
Unlike the usual convention in critical dynamics, we have explicitly
expressed the temperature $T$, which will be useful below.
The correlation function of the noise ensures that the probability that
an order parameter configuration $\psi$ will occur is $P \{ \psi \} =
Z^{-1} \exp [- F \{ \psi \} / T ]$ (where $Z = \int D \{ \psi \} \exp
[- F \{ \psi \} / T ])$.

The main object we consider within this model is the operator of heat
current (\ref{heat-current}). This form may be understood in terms of
the transport of free energy. In Sec.~\ref{Tgradient} we note how it
arises from considerations involving the application of a temperature
gradient to the system. It also arises in derivations of time-dependent
Ginzburg-Landau equations from microscopics. Note that the heat current
operator is a dynamic object involving a time derivative (unlike the
electric current); thus its properties are dependent on the choice of
dynamics.

When particle-hole symmetry is broken, the relaxation rate $\Gamma_0$
becomes complex, and model A admits non-dissipative, or reactive,
terms. While we expect such terms not to contribute to heat
transport, the question of incorporating this
expectation into the critical dynamics treatment of the heat current is
not fully resolved. In this paper, we therefore proceed with the
assumption particle-hole symmetry.

\section{Scaling dimension of $\mathbf{j}^Q$}
\label{dimension}

In this section, we consider the scaling dimension $d_Q$ of the heat
current operator. The dimension $d_Q$ will then allow us to write
scaling forms for properties involving the heat current, as we do in
Sec.~\ref{transport} for thermal and thermoelectric transport. Naively,
one may expect the emergence of an anomalous dimension for this
operator (i.e., a deviation from its Gaussian value). In particular,
because of the explicit time derivative in $\mathbf{j}^Q$
[Eq.~(\ref{heat-current})], one may expect the anomalous dimension to
involve the dynamical critical exponent $z$. In this section we show
that these expectations are false; we find $d_Q = d + 1$ in $d$
dimensions, with no anomalous dimension. We do so first by an exact
argument, then by sketching the calculation of $d_Q$ to second order in
an $\epsilon$ expansion.

We begin by replacing the dynamic stochastic equation with an effective
action for calculating dynamic correlators~\cite{Zinn-Justin}. Consider
the Langevin equation given by Eq.~(\ref{model-A}) (in this section we
set $T = 1$). The noise term has a Gaussian distribution,
\begin{equation}
P \{ \zeta (x, t) \} \propto \exp \left( - \int d^d x \, dt \,\frac{
|\zeta (x, t)|^2}{2 \Gamma_0} \right) .
\end{equation}
Enforcing Eq.~(\ref{model-A}) by a $\delta$ function, expectation
values, when averaged over the noise distribution, take the form
\begin{equation}
\langle \cdots \rangle_\zeta \propto \int D \psi \, D \zeta \, \cdots
\, \delta \left( \frac{\partial \psi}{\partial t} + \Gamma_0
\frac{\delta F}{\delta \psi^*} - \zeta \right) P \{ \zeta (x, t) \} .
\end{equation}
The integral over the noise may now be performed, yielding an effective
action for $\psi$,
\begin{equation}
\langle \cdots \rangle \propto \int D \psi \, \cdots \, e^{-
S_{\text{eff}} \{ \psi \} } ,
\end{equation}
where
\begin{eqnarray}
S_{\text{eff}} \{ \psi \} & = & \frac{1}{2 \Gamma_0} \int d^d x \, dt
\, \left| \frac{\partial \psi}{\partial t} + \Gamma_0 \frac{\delta
F}{\delta \psi^*} \right|^2 \\ \nonumber & = & \frac{1}{2 \Gamma_0}
\int d^d x \, dt \, \left| \frac{\partial  \psi}{\partial t} \right|^2
+ \left| \Gamma_0 \frac{\delta F}{\delta \psi^*} \right|^2 .
\end{eqnarray}
The last step uses the fact that the cross term in the action is a full
derivative and hence vanishes upon integration. In the above, we have
ignored the Jacobian term, which is irrelevant for the renormalization
group analysis~\cite{Zinn-Justin}. The problem of calculating dynamical
correlation function is thus expressed in terms of a new effective
action $S_{\text{eff}}$, which may be viewed either as a quantum
mechanical action (the view adopted here), or equivalently as a
classical statistical mechanical action in $d+1$ dimensions.

The exact argument regarding the dimension of the heat current operator
is  based on the following observation: The heat current,
Eq.~(\ref{heat-current}), is proportional to the conserved momentum
density of the effective Lagrangian,
\begin{equation}
T_{0i} = \frac{\partial \mathcal{L}_{\text{eff}}}{\partial (\partial_0
\psi)} \partial_i \psi + \text{c.c.} = \frac{1}{\Gamma_0} \left(
\frac{\partial \psi^*}{\partial t} \nabla_i \psi + \text{c.c.} \right) .
\end{equation}
This observation essentially sets the dimension of the heat current
operator; it does not have an anomalous dimension because it is
conserved. A similar argument, but in the purely static theory,
fixes the dimension of the electric current operator as $d_e = d-1$.

A standard argument turns the conservation law into a Ward identity.
Consider a time-ordered correlation function of $T_{0i}$ with a set of
operators $\psi$ (whose dimension is known),
\begin{equation}
\left\langle \mathcal{T} T^{\mu i} (x) \psi (x_1) \ldots \psi (x_n)
\right\rangle .
\end{equation}
(Here, we use $x$ to denote both $x$ and $t$ in the arguments of the
fields.) Using $\partial^\mu T_{\mu i} = 0$ (as $T_{\mu i}$ is a
conserved current  for each of its components $i$), we only need to
account for the time-derivative on the step functions due to
time-ordering,
\begin{eqnarray}\label{exact-arg}
\lefteqn{\partial^\mu \left\langle \mathcal{T} T_{\mu i} (x) \psi (x_1)
\ldots \psi (x_n) \right\rangle} \\
\nonumber & & = \sum_{j = 1}^n
\delta (t - t_i) \langle \psi (x_1)  \ldots [T_{0 i} (x), \psi (x_j)]
\ldots \psi (x_n) \rangle  \\
\nonumber & & = -i \sum_{j = 1}^n \delta^{(d +
1)} (x - x_i) \langle \psi (x_1) \ldots \partial_i \psi(x_j) \ldots
\psi (x_n) \rangle .
\end{eqnarray}
The dimension of the heat current operator, $d_Q = d + 1$, may be
deduced from Eq.~(\ref{exact-arg}) by comparing the term with $\mu = 0$
in the first line (involving the required heat current operator) with
the last line of the equation (where all the dimensionalities are
known). The dimensionality of the fields $\psi (x_i)$ cancel between
the two sides, as does the factor $z$ between the time derivative in
the first line and the time component of the $\delta$ function in the
last line. The spatial part of the $\delta$ function and the additional
spatial derivative in the last line give the dimension of the heat
current operator.

As the result above is not very transparent in terms of its physical
content it is instructive to see how this works in an expansion about
the Gaussian fixed point; indeed, this is how we came across this
result in the first place. Specifically we consider the expansion in
$\epsilon = 4 - d$ to second order (the first non-trivial order) for
the dimension of the heat current. We find that the coefficients of
this expansion vanish, thus corroborating the exact argument above.

The calculation proceeds by adapting the method for computing
dimensions of composite operators outlined by Wilson and
Kogut~\cite{Wilson-Kogut} to the dynamic theory. For convenience, the
complex order parameter $\psi$ is represented in this calculation in
terms of its real and imaginary components, $\psi = \phi_1 + i \phi_2$,
and the heat current operator is then
\begin{equation}
\mathbf{j}^Q \propto - \frac{\partial \phi_1}{\partial t} \, \nabla
\phi_1 -  \frac{\partial \phi_2}{\partial t} \, \nabla \phi_2 .
\end{equation}
We consider the following correlation function involving the heat
current operator
\begin{eqnarray}\label{U}
\lefteqn{\mathbf{U} (\mathbf{q}, \omega) = \int d \mathbf{x} \, d
\mathbf{x'} \, d t \, d t' e^{i \mathbf{q} \cdot (\mathbf{x} -
\mathbf{x'}) - i \omega (t - t')}} \\ \nonumber & & \qquad \qquad
\qquad \qquad \times \, \left\langle \phi_i (\mathbf{x}, t) \phi_i
(\mathbf{x'}, t') \mathbf{j}^Q (0, 0) \right\rangle .
\end{eqnarray}
Following Wilson and Kogut, the correlation function has been chosen
both for its simplicity and because it avoids the mixing in of
operators involving total derivatives.

By simple counting of dimensions, the correlation function may be
expressed (in terms of an unknown scaling function $\mathbf{F}$) as
\begin{equation}\label{Uxi}
\mathbf{U} = \xi^{2d + 2z - 2 d_s - d_Q} \mathbf{F} (\mathbf{q} \xi,
\omega \xi^z) ,
\end{equation}
Here, $\xi$ is the correlation length, $d_Q$ is the unknown dimension
of the heat current operator, and $d_s = (d - 2 + \eta) / 2$ is the
dimension of the operator $\phi_i$.
This result may be expressed in terms of the $\langle \phi_i \phi_i
\rangle$ static correlator at $\mathbf{q} = 0$, $r^{-1} \propto \xi^{2
- \eta}$. Expanding to linear order in $\mathbf{q}$ and $\omega$, we
thus have
\begin{equation}\label{U-power-r}
\mathbf{U} \propto \mathbf{q} \, \omega \, r^{(d_Q + \eta - d - 3 - 3z)
/ (2 - \eta)}.
\end{equation}
This result will be compared with a perturbative calculation of
$\mathbf{U}$ to obtain the result for $d_Q$.

\begin{figure}
\includegraphics[width=3.25in]{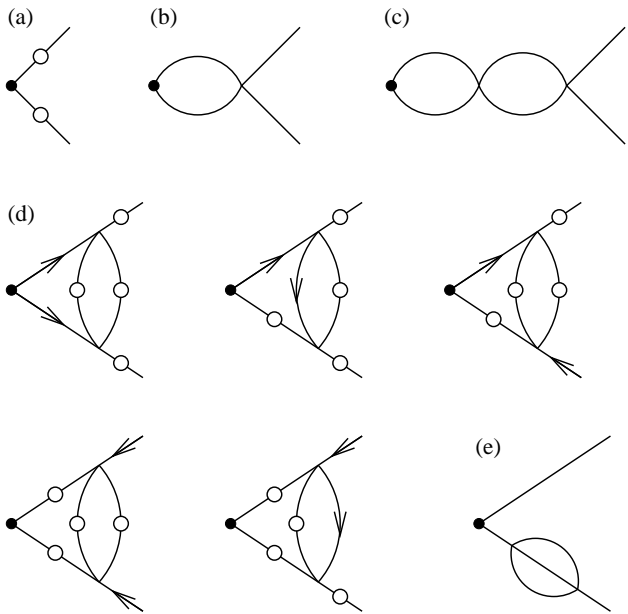}
\caption{\label{fig} Diagrams appearing in the correlator $\mathbf{U}$
to second order in $u$. The heat current vertex $\mathbf{j}^Q$ is
denoted by a black circle. A line with an arrow denotes the bare
propagator $G_0$, and a line with an empty circle denotes the bare
correlator $C_0$ [in (b), (c), and (e) the diagrams are shown without
presenting all possible arrangements of propagator and correlator
lines]. Diagram (a) is the bare (or zeroth order) contribution.
Diagrams (b) and (c) are first and second order diagrams, respectively,
which have a vanishing contribution. Diagrams which contribute to
second order are given in (d) and (e).
}
\end{figure}

We briefly outline the calculation of $\mathbf{U}$. The perturbative
method for the dynamic case (see, e.g., Ref.~\cite{Ma}) begins by
presenting the fields as
\begin{eqnarray}\label{phi-purb}
\lefteqn{\phi_i (\mathbf{k}, \omega) = \phi_i^{(0)} (\mathbf{k}, \omega) -
u G_0 (\mathbf{k}, \omega) \sum_j \int  \frac{d \mathbf{k'} \, d \mathbf{k''}
\, d \omega' \, d \omega''}{(2 \pi)^{2 (d + 1)}}} \nonumber \\
& & \!\!\!\!\!\!\! \times \, \phi_j (\mathbf{k'}, \omega') \phi_j
(\mathbf{k''}, \omega'') \phi_i (\mathbf{k} - \mathbf{k'} -
\mathbf{k''}, \omega - \omega' - \omega''). \qquad \;
\end{eqnarray}
Here, $\phi_i^{(0)} = G_0 (\mathbf{k}, \omega) \zeta_i (\mathbf{k},
\omega) / \Gamma$ is the free field, and
\begin{equation}
G_0 (\mathbf{k}, \omega) = \frac{1}{r + k^2 - i \omega / \Gamma}
\end{equation}
is the free propagator. Equation~(\ref{phi-purb}) is then used to
generate the  different diagrammatic contributions to the desired order
in $u$. In contrast with the static perturbation theory, such diagrams
contain lines corresponding to both the propagator $G_0$ and the
correlator
\begin{eqnarray}
\lefteqn{\left\langle \phi_i^{(0)} (\mathbf{k}, \omega) \phi_j^{(0)}
(\mathbf{k'}, \omega') \right\rangle =} \\
\nonumber & & \qquad \qquad (2 \pi)^{d + 1} \delta (\omega + \omega')
\delta (\mathbf{k} + \mathbf{k'}) \delta_{ij} C_0 (\mathbf{k}, \omega)
,
\end{eqnarray}
where $C_0 (\mathbf{k}, \omega) = (2 / \omega) \text{Im} G_0
(\mathbf{k, \omega})$. The diagrams which arise to second order in $u$
are presented in Fig.~\ref{fig}.

At Gaussian order [see Fig.~\ref{fig}(a)] and to linear order in
$\mathbf{q}$ and $\omega$, we have
\begin{equation}
\mathbf{U}_0 = \frac{4 \mathbf{q} \, \omega}{\Gamma^2 r^4} .
\end{equation}
To first order in $u$ we have the diagram in Fig.~\ref{fig}(b), but its
contribution vanishes due to symmetry considerations [as does the
second order diagram in Fig.~\ref{fig}(c)]. (Diagrams in which a closed
loop appears on one of the propagators are included by the
renormalization of $r$.) To second order in $u$, we have the diagrams
in Fig.~\ref{fig}(c)--(e). Of these, we have illustrated in
Fig.~\ref{fig}(d) the fact that each of these diagrams is in fact a set
of diagrams with different placements of propagator and correlator
lines. Note that the external legs in the different diagrams may be
either propagator or correlator lines. For this reason, in contrast
with the static case, we calculate all diagrams appearing to a given
order, not only those which are irreducible with respect to the
external legs.

In the calculation of the diagrams, we are interested in the infrared
divergence when $r \rightarrow 0$. The diagrams are also ultraviolet
divergent, which we regularize by dimensional regularization  (see.,
e.g., Ref.~\cite{Binney-etal}).
The actual calculation of these diagrams is rather lengthy, and we will
spare the reader the details. Very briefly, the various diagrams are
evaluated using the value of $u$ at the fixed point, which to order
$\epsilon$ is $u^* = \pi^2 \epsilon / 5$. It is convenient to express
the result in terms of the exponent $\eta$ (to second order in
$\epsilon$, $\eta = \epsilon^2 / 50$) and the coefficient $c = 6 \ln
\frac{4}{3} - 1$, which appears in the result for the dynamical
critical exponent,  $z = 2 + c \eta$~\cite{Halperin-Hohenberg-Ma}. The
result is
\begin{equation}\label{UU-power-r}
\mathbf{U} = \mathbf{U}_0 \left( 1 - \frac{3}{2} (c + 1) \eta \ln r
\right) \propto r^{-4 - 3(c + 1) \eta / 2}.
\end{equation}
A comparison of Eqs.~(\ref{U-power-r}) and~(\ref{UU-power-r}) gives
$d_Q = d + 1$ to second order in $\epsilon$. We thus do not find an
anomalous dimension for the heat current operator in the $\epsilon$
expansion, in agreement with the general argument given above.

\section{Introducing a temperature gradient}
\label{Tgradient}

In previous sections, the temperature~$T$ was constant throughout the
sample. A constant temperature may be simply absorbed by an appropriate
rescaling of variables, so that it does not appear explicitly in the
problem. Indeed, by rescaling the order parameter, free energy, and
noise using $\psi \rightarrow \sqrt{T} \psi$, $F \rightarrow T F$, and
$\zeta \rightarrow \zeta / \sqrt{T}$,
Eqs.~(\ref{free-energy})--(\ref{noise}) may be rewritten without an
explicit temperature (this also requires rescaling the coefficient of
the quartic term in the free energy). This is the form traditionally
used in the literature on critical dynamics~\cite{Hohenberg-Halperin}.

For thermal and thermoelectric transport, however, we should also
consider the application of temperature gradients in the sample.
For this purpose, we have retained the explicit temperature
dependence of the noise correlator, Eq.~(\ref{noise}). Other
parameters of the problem may also have an implicit temperature
dependence; however, at least for linear response to a temperature
gradient, the dependence of parameters on temperature is not
important for transport and is subsequently ignored.  The reason
for this is that if we ignore the temperature dependence in the
noise, the system then has spatially-dependent local couplings
(through their temperature dependence) but remains in isothermal
equilibrium; thus it carries no transport currents.

Rescaling Eq.~(\ref{model-A}), as discussed above, leads to the
following equation of motion, to linear order in the temperature
gradient,
\begin{eqnarray}\label{nablaT}
\lefteqn{\frac{\partial \psi}{\partial t} =} \\ \nonumber & & - \,
\Gamma_0 \left[ r_0 \psi + 2 u_0
|\psi|^2 \psi - \nabla^2 \psi - (\nabla \psi) \cdot (\nabla T) / T
\right] + \zeta .
\end{eqnarray}
The dependence of the correlator (\ref{noise}) on temperature has
been exchanged for a term proportional to $\nabla T$ in the
stochastic equation. The equation may still be recast in the form
(\ref{model-A}), provided an additional term, of the form $-
\mathbf{j}^Q \cdot \tilde{\mathbf{A}}$, is added to (the integrand
of) the free energy, Eq.~(\ref{free-energy}). Here, $\mathbf{j}^Q$
is the heat current, Eq.~(\ref{heat-current}). It is coupled to a
``gravitational'' vector potential $\mathbf{\tilde A}$. The notion
of a gravitational field was introduced by
Luttinger~\cite{Luttinger} as a fictitious device for studying
temperature gradients (see also
Ref.~\cite{Cooper-Halperin-Ruzin}). Here, we briefly discuss this
point, noting the analogy between the treatment of electric and
thermal currents. In particular, Einstein relations for thermal
currents relate the response to $(\nabla T) / T$ with the response
to a gravitational force $\tilde \mathbf{E} = - \nabla \tilde
\phi$, where $\tilde \phi$ is Luttinger's gravitational field.
By performing a gravitational
gauge transformation, the gravitational force is expressed in
terms of a gravitational vector potential, $\tilde{\mathbf{E}} = -
\partial \tilde{\mathbf{A}} / \partial t$. To linear order in the
gravitational force, the gravitational vector potential
$\tilde{\mathbf{A}}$ is then coupled to $\mathbf{j}^Q$ in the free
energy of model A. 
The equation of motion~(\ref{model-A}), coupled with the
Einstein relations, reproduces Eq.~(\ref{nablaT}).

This understanding of how a thermal gradient enters the free energy is
useful on several fronts. First, it verifies the form of the heat
current operator, Eq.~(\ref{heat-current}). Second, it replaces
temperature gradients with gravitational fields. Luttinger's original
motivation was to express thermal transport coefficients in terms of
the corresponding correlation functions. In the next section, we follow
a similar strategy for writing scaling forms for the thermal current.

\section{Thermal transport}
\label{transport}

In this section, we consider the critical behavior of thermal transport
coefficients in model A dynamics. Using the result for the scaling
dimension of the heat current from Sec.~\ref{dimension}, $d_Q = d + 1$,
we write the general scaling relations for the electric and heat
currents, in presence of an electric, magnetic, and gravitational
fields,
\begin{eqnarray}
\mathbf{j}^e & = & \xi^{-d+1} \mathbf{f}^e \left[ E \xi^{1 + z}, B
\xi^2, (\nabla \tilde{\phi}) \xi^{-1 + z} \right] \\
\mathbf{j}^Q & = & \xi^{-d-1} \mathbf{f}^Q \left[ E \xi^{1 + z}, B
\xi^2, (\nabla \tilde{\phi}) \xi^{-1 + z} \right] .
\end{eqnarray}
These equations are extensions of known results for electric transport
(see, e.g., Ref.~\cite{Fisher-Fisher-Huse}).

We will use these scaling relations here to deduce the critical
behavior of various linear response coefficients. We note that the
assumption of particle-hole symmetry implies that $\sigma_{xy} =
\alpha_{xx} = \kappa_{xy} = 0$ (it is necessary to break particle-hole
symmetry to discuss the critical properties of these transport
coefficients). For the conductivity, as well as for the magnetic
susceptibility, the known results~\cite{Fisher-Fisher-Huse} may be
confirmed,
\begin{equation}
\sigma_{xx} \propto \xi^{2 + z - d}, \qquad \chi \propto \xi^{4 - d} .
\end{equation}
Our result for the heat current operator now enables us to make
analogous statements for transport coefficients involving the heat
current operator, $\alpha_{xy}$ and $\kappa_{xx}$.

We begin with the transverse thermoelectric response $\alpha_{xy}$,
which is the coefficient of interest in studying the Nernst effect.
More precisely, we will consider the response to linear order in the
magnetic field. This may be calculated either from the heat current
response to an applied electric and magnetic field, or as the electric
current response to the applied gravitational and magnetic fields. In
both cases, the current which is obtained is the \emph{total} current,
from which magnetization contributions, which do not contribute to
transport, must be subtracted~\cite{Cooper-Halperin-Ruzin}. For the
total currents, we find
\begin{equation}
\frac{j_{\text{tot}}^Q}{E B} \propto \frac{j_{\text{tot}}^e}{(\nabla
\tilde \phi) B} \propto \xi^{2 + z - d}.
\end{equation}
The magnetization component of these total currents is given
by~\cite{Cooper-Halperin-Ruzin}
\begin{equation}
j_{\text{mag}}^Q = \mathbf{M} \times \mathbf{E} , \qquad
j_{\text{mag}}^e = - \mathbf{M} \times \nabla \tilde \phi .
\end{equation}
Using the result for the magnetic susceptibility above, we see that the
situation is different dependent upon the value of the critical
dynamical exponent.

For $d \geq 4$ mean-field exponents hold, and the dynamical critical
exponent is $z = 2$. In this case, the total and magnetization pieces
of the current diverge with identical power laws (cf.\
Ref.~\cite{Ussishkin-Sondhi-Huse}). The difference of two terms with
identical singularities may be of the same singular behavior or a
weaker one. Thus, the critical dynamics analysis presented here does
not determine conclusively the power of $\alpha_{xy}$ when $z = 2$. For
Gaussian fluctuations, $\alpha_{xy}$ does have the same singularity as
the total currents and magnetization contributions, $\alpha_{xy} / B
\propto \xi^{4-d}$ ~\cite{Ussishkin-Sondhi-Huse}. It is quite possible
that the same occurs also in two dimension (where $z = 2$ as well),
suggesting that $\alpha_{xy} / B \propto \xi^2$ above the
Kosterlitz-Thouless transition temperature. However, a separate
analysis is required for this case which we defer to a future work.

For three dimensions, which corresponds to the actual transition in the
cuprates due to interlayer coupling, the analysis of
Ref.~\cite{Halperin-Hohenberg-Ma} gives $z = 2 + c \eta > 2$. In this
case, the magnetization currents are less singular than the total
currents from which they are subtracted, and  hence the result for
$\alpha_{xy}$ to linear order in $B$,
\begin{equation}\label{alphaxy}
\alpha_{xy} / B \propto \xi^{2 + z - d},
\end{equation}
becomes unproblematic \cite{fn-nonvanish}. We note that the
dimensionless ratio $T \alpha_{xy} / M$ (where $M = \chi B$ is the
magnetization) diverges at the critical point as $\xi^{z-2}$.

Finally, we consider the thermal conductivity, for which we find
\begin{equation}
\kappa_{xx} \propto \xi^{-2 + z - d}.
\end{equation}
In particular, the thermal conductivity will have a singular, but
non-diverging contribution in model A in three dimensions. In contrast,
Vishveshwara and Fisher argued in a recent paper~\cite{Vishveshwara}
that the thermal conductivity is analytic at the critical point using a
model C formulation (which involves a conserved energy density in
addition to a non-conserved order parameter). There are several things
to note about this difference in the critical behavior obtained in the
two works. First, the coupling of the energy variable to the order
parameter is known to be irrelevant (as $\alpha<0$ with $n=2$
components of the order parameter), and we expect the critical behavior
of model C to be equivalent to that of model A~\cite{HHM}. 
Second, we note that
singular behavior is already obtained for the thermal conductivity for
Gaussian fluctuations in model A~\cite{Ussishkin-Sondhi-Huse}, a result
which may also be obtained from microscopics (which are energy
conserving). Finally, it seems to us that in Ref.~\cite{Vishveshwara},
a treatment of the order parameter contribution to the heat current is
absent.

\section{Final comments}
\label{summary}

Our results for the critical behavior of the transverse thermoelectric
response $\alpha_{xy}$ and the thermal conductivity $\kappa_{xx}$ are
based on the observation that the heat current operator has no
anomalous dimension. It is worth noting that both electric and heat
currents share this property; in particular they are independent of the
dynamical critical exponent $z$. Readers familiar with the lore
of quantum critical phenomena should note that conserved currents
{\it do} exhibit $z$ in their scaling dimensions at critical points;
it is their corresponding densities which do not. Thus the meaning
of the phrase ``absence of anomalous dimension'' is different in
these two cases.

We briefly comment on the possibility of observing our results in
experiment. In the moderately two dimensional cuprates, such as
YBCO, there is evidence for a relatively large 3D-XY critical
regime. In this regime our results for  $\alpha_{xy}$, to linear
order in the field, at temperatures approaching $T_c$ can be
tested via measurements of the Nernst effect and the conductivity.
Another possibility, instead of attempting to extract critical
exponents, is to consider the dimensionless ratio $T \alpha_{xy} /
M$. This ratio equals $1/2$ in the region of Gaussian
fluctuations~\cite{Ussishkin-Sondhi-Huse}. As the temperature is
decreased towards $T_c$, the value of this ratio increases,
eventually diverging as $\xi^{z-2}$. In the highly two dimensional
cuprates such as BSSCO the scaling theory is on somewhat weaker
grounds due to the degeneracy between the dimensions of the total
current and the magnetization. However, absent an exact
cancellation, it should still correctly predict that $\alpha_{xy}$
to linear order in $B$ diverges with the same critical exponent
(i.e. $\sim\xi^2$) as the magnetization as the Kosterlitz-Thouless
transition is approached (except perhaps for a confluent logarithm
at the lower critical dimension.)

In summary, we considered the critical scaling of thermal transport
near the finite temperature transition to the superconducting state,
within the framework of critical model A dynamics. This leads to
specific predictions for the critical exponents of thermal transport
coefficients, and in particular for the Nernst effect.

\begin{acknowledgments}
We thank Tom Lubensky for illuminating discussions and valuable
input. This work was supported by the NSF through
MRSEC grant DMR-02-13706 at Princeton, by the David and Lucile
Packard Foundation (SLS), and by NSF grant EIA-02-10736 (IU).

\end{acknowledgments}


\begin{thebibliography}{5}
\bibitem{alpha-xy}
The Nernst signal is the transverse electric field response to a
temperature gradient under open circuit conditions (Hall thermopower).
It is related to the conductivity tensor $\sigma$ and the
thermoelectric tensor $\alpha$ using
$$
\frac{E_y}{(- \nabla T)_x} = \frac{\alpha_{xy} \sigma_{xx} -
\alpha_{xx} \sigma_{xy}}{\sigma_{xx}^2 + \sigma_{xy}^2} \approx
\frac{\alpha_{xy}}{\sigma_{xx}} ,
$$
where the last approximation is appropriate when the term
containing $\alpha_{xy}$ dominates the Nernst signal and the Hall
angle is small.
\bibitem{fn-model}
Of course, in a real superconductor both electrical charge and
energy are conserved, and a reader might wonder if model A
dynamics without any conservation laws can really capture the
correct critical dynamics. This is a traditional assumption that
we use here.
The heuristic justification for ignoring charge conservation
is that the Coulomb interaction makes the corresponding mode (the
plasmon) gapped, so it does not enter in the hydrodynamics.
Including energy conservation would
"promote" model A to model C. However, the coupling
of the energy density to the order parameter in model C
is irrelevant (in both two and three dimensions), and the two models
should result in the same critical behavior (see Ref.~\cite{HHM}).
\bibitem{fn-nonvanish}
Of course we assume that the coefficents of the leading power laws
do not fortuitously vanish. As they do not do so at the Gaussian
fixed point, this seems a safe assumption.
\end{thebibliography}
\end{document}